\documentclass[preprint,aps,prd,amsfonts,amsmath,amssymb,nofootinbib,longbibliography]{revtex4-1}

\usepackage{ifpdf}
\usepackage{braket}

\ifpdf
    \usepackage[pdftex,colorlinks=true,plainpages=false,pdfpagelabels]{hyperref}
    \usepackage[pdftex]{color}
    \hypersetup{pdfauthor = {John Kehayias}}
    \hypersetup{pdftitle = {Generalized Gaugino Condensation in Super Yang-Mills Theories: Discrete R-Symmetries and Vacua}}
    \hypersetup{colorlinks = true}
\else
    \usepackage[colorlinks=true]{hyperref}
\fi

\newcommand{\bbar}[1]{\overline{#1}}  
\def\Tr{\mathrm{Tr~}}

\begin{document}

\title{Generalized Gaugino Condensation in Super Yang-Mills
  Theories:\\ Discrete R-Symmetries and Vacua}

\author{John Kehayias} \email{kehayias@physics.ucsc.edu}
\affiliation{Santa Cruz Institute for Particle Physics and Department
  of Physics,\\ University of California, Santa Cruz, CA 95064, USA}

\date{\today}

\begin{abstract}
  \noindent One can define generalized models of gaugino condensation
  as theories which dynamically break a discrete R-symmetry, but do
  not break supersymmetry.  We consider general examples consisting of
  gauge and matter fields, and the minimal number of gauge singlet
  fields to avoid flat directions in the potential.  We explore which
  R-symmetries can arise, and their spontaneous breaking.  In general,
  we find that the discrete symmetry is $\mathbb{Z}_{2b_0R}$ and the
  number of supersymmetric vacua is $b_0$, where $b_0$ is the
  coefficient of the one-loop beta function.  Results are presented
  for various groups, including $SU(N_c), SO(N_c), Sp(2N_c)$, and
  $G_2$, for various numbers of flavors, $N_f$, by several methods.
  This analysis can also apply to the other exceptional groups, and
  thus all simple Lie groups.  We also comment on model building
  applications where a discrete R-symmetry, broken by the singlet
  vevs, can account for $\mu$-type terms and allow a realistic Higgs
  spectrum naturally.

\end{abstract}

\maketitle

\section{Introduction}

Gaugino condensation \cite{Taylor:1982bp, *Veneziano:1982ah} is
frequently discussed in considering problems of supersymmetry dynamics
and model building.  There are several distinguishing features of this
non-perturbative effect: (i) it does not break supersymmetry, (ii) it
breaks a discrete R-symmetry, and (iii) it generates a scale
dynamically (used, for instance, in the ``retrofitting'' procedure of
\cite{retro}).  In \cite{disr_paper} a large class of theories with
gauge, matter, and gauge singlet fields with these features was
introduced.  This is a generalization of gaugino condensation,
possessing the above properties.  In particular, \cite{disr_paper}
explored the significant model building consequences of the R-symmetry
being broken by an order parameter with mass dimension less than
three.


In this work we explore theories with generalized gaugino condensation
in more detail, and for a wide range of gauge groups.  This extends
and generalizes the supersymmetric $SU(N_c)$ theory used in
\cite{disr_paper}.  We find quite generally that the discrete symmetry
and number of supersymmetric vacua are counted by the one-loop beta
function coefficient, $b_0$.  The discrete R-symmetry is found by considering
the R-charge of an instanton in the theory with a continuous
R-symmetry at the classical level.  We find that requiring that the
instanton be uncharged under a discrete subgroup gives
$\mathbb{Z}_{2b_0R}$.  This calculation method is even more general
than finding the number of vacua, since the instanton calculation is
done without assuming the group or even the representations of the
matter content.

In all of the theories we study, the R-symmetry is broken to a
$\mathbb{Z}_2$ by gaugino condensation and the singlet vevs, leading
to a discrete set of supersymmetric vacua.  The dynamics responsible
for breaking the symmetry depends on the coupling regime.  In some
regions the vacua can be calculated in general by utilizing a generic
effective superpotential, motivated by (and matching) several known
examples, or by integrating out heavy quarks.  In other regions of a
particular theory, the specific low energy dynamics are important.  We
find the number of vacua to be $b_0$, matching what one expects from
the breaking of the R-symmetry.  These results are shown explicitly in
several examples with different gauge groups, and it seems possible
that this holds, in at least some regions, for any simple Lie group.

The paper is organized as follows: in the next section we derive
general results for the discrete R-symmetry and number of vacua.  In
Section \ref{sec:su} these results are derived in different regions of
the parameter space for $SU(N_c)$ models, for a wide range of $N_f$,
the number of flavors.  Depending on the value of $N_f$, a few
different methods are used.  Section \ref{sec:sosp} shows that these
results hold for $SO(N_c), Sp(2N_c),$ and $G_2$ models.  We also
comment on the difficulties with the exceptional groups, and why it is
possible these results may apply here as well.  This would then
encompass all the simple Lie groups.  This work also extends the model
building introduced in \cite{disr_paper}, which we comment on
(including using a recent NMSSM-like model) in Section
\ref{sec:model}.  A brief discussion and concluding remarks follow
that.

\section{General Results}\label{sec:gen}

Many of the topics and techniques used in this work are well
summarized in the review by Peskin \cite{dualityreview}.  Let us first
define some relevant quantities and conventions before the general
calculations.

$N_c$ will denote the number of colors (i.e.~as in $SU(N_c)$),
although some care must be taken with factors of $2$ for the
symplectic group.  The quark superfields are $Q_i$ and $\bbar{Q}_i$,
which will be in the fundamental or anti-fundamental representation of
the group, as appropriate for the given gauge group (in the vector
representation in the case of $SO(N_c)$).  However, for the
general calculation of the discrete R-symmetry, the quarks may be in
any representation.  We define the (gauge invariant) ``meson''
superfield $M_{ij} = Q_i\bbar{Q}_j$, where $i$ and $j$ run from $1$ to
$N_f$.  For our purposes it suffices to take derivatives of the
superpotential $W$ with respect to $M$, but one can also work directly
with the $Q$'s.  The gauge singlets will be denoted by $S_{ij}$ where
$i$ and $j$ again run from $1$ to $N_f$ (up to a $2$ for $Sp$);
there are $N_f^2$ singlets, corresponding to each of the possible
flavor combinations.

Although we will use gauge singlets throughout this work, one could
use some other representation of the gauge group.  We rely only on a
cubic self-coupling for these fields and no dimensionful couplings in
the superpotential, so fields in the adjoint could also work, for
instance.  In a practical sense, choosing a different representation
may be of use in model building, or further constrained by other
requirements.

The gauginos are represented by $\lambda$ (in general, spinor indices
will be dropped).  We will also assume flavor-symmetric solutions and
single coupling constants (rather than a matrix in flavor space).

Our normalization for the group theory constants $C(r_i)\delta_{ab} =
\Tr \{t_a t_b\}_{r_i}$, where $r_i$ denotes the representation, is
such that the fundamental representation has $C = 1/2$.  Then, for
instance, the adjoint of $SU(N_c)$ has $C = N_c$.

\subsection{The Discrete R-Symmetry}
To find the (non-anomalous) discrete R-symmetry, we start with a
continuous $U(1)_{R}$.  The instanton breaks this symmetry, and we look
at what discrete subgroup can remain (i.e.~the instanton is not
charged under the subgroup).

First, let's define the R-charges of the superfields, where our
superpotential will have interactions of the form $SQ\bbar{Q} + S^3$.
We will start by defining the gaugino transformation parameter,
$\beta$:
\begin{equation}
  \lambda \to \beta\lambda.
\end{equation}
The requirement, familiar in the case of a continuous R-symmetry, that
$W$ (and $W_\alpha^2$) must have R-charge $2$ becomes a transformation
by an overall factor of $\beta^2$.  In other words, in this notation
the actual R-charge is given by powers of $\beta$.  Gaugino
condensation, $\braket{\lambda\lambda}$, and a cubic self-interaction
of the singlets (in $W$) both have the same total R-charge.  The $S$'s
must transform as follows:
\begin{equation}
  S_{ij} \to \beta^{2/3}S_{ij}.
\end{equation}
Finally, the quark superfields will be coupled to the singlets as
$QS\bbar{Q}$, which also must transform with a factor of $\beta^2$.
$Q_i$ and $\bbar{Q}_i$ have the same R-charge, and the fermionic
component differs by a power of $\beta$ from this (due to the charge
of the Grassmann $\theta$ coordinate, which transforms like
$\lambda$)\footnote{We note that these R-charges are the same as the
  usual non-anomalous R-charges for $SU(N_c), SO(N_c),$ and $Sp(2N_c)$
  when taking into account the discrete arithmetic of the group we
  find below.}:
\begin{align}
  Q_i &\to \beta^{2/3}Q_i\\
  \psi_Q &\to \beta^{-1/3}\psi_Q.
\end{align}
The interaction terms mentioned above will be made more explicit in
the later sections.

To determine the R-charge of the instanton, we only need to know how
many fermions are involved.  The number of zero modes for a fermion in
the representation $r_i$ appearing in the instanton is given by twice
the group theoretical coefficient defined previously:
\begin{equation}
  n_i = 2C(r_i).
\end{equation}

The calculation is now very simple: we have the gauginos in the
adjoint representation (denoted $A$), and quarks in the fundamental or
anti-fundamental representation\footnote{Actually, nothing here
  depends on the representation of the quarks; the calculation is
  general.  In the following sections, however, we will have the
  quarks in the fundamental or anti-fundamental representations.}
(denoted as just $r_i$).  The total R-charge (power of $\beta$) is
then:
\begin{equation}
  2C(A) - \frac{1}{3}\sum_i 2C(r_i) = \frac{2}{3}(3C_2(A) - \sum_i C(r_i))
\end{equation}
where we used the fact that in the adjoint, $C(A) = C_2(A)$, where
$C_2(A)$ is the quadratic Casimir operator of the adjoint
representation.  We recognize this as being proportional to the
coefficient of the one-loop beta function:
\begin{equation}
  b_0  = 3C_2(A) - \sum_i C(r_i)
\end{equation}
Therefore the discrete subgroup of the $U(1)_{R}$ that is left over by
the instanton is $\mathbb{Z}_{2b_0 R}$.  Since this will ultimately be
broken down to just a $\mathbb{Z}_2$ by gaugino condensation and the
singlet vev, we expect to see $b_0$ vacuum states, which we will show
in the next section.

\subsection{The Number of Vacua}
We will calculate the number of vacua by considering a general
superpotential, including an effective superpotential term.  This
analysis is only for when such a term exists, but we will see how to
extend this result quite generally.

The interaction terms in the general superpotential from including the
singlets, $S_{ij}$, are
\begin{equation}
  W_S = yS_{ij}M_{ij} + \frac{\gamma}{3}\Tr S^3,
\end{equation}
with $y$ and $\gamma$ coupling constants (for simplicity, we do not
write them as more general matrices).

Let's consider an effective superpotential term of a generic form, which
can incorporate the known effective superpotential term, when it
exists, from $SU(N_c), SO(N_c), Sp(2N_c)$ and $G_2$ supersymmetric
gauge theories.  The ingredients are the energy scale of the theory,
$\Lambda$, which has a power determined by the beta function, the
meson superfield, which we take to just be some power of the matrix
elements (this is easy to see in e.g.~flavor symmetric solutions of
$SU(N_c)$), and the fact that the total mass dimension must be $3$:
\begin{equation}\label{eq:weff_gen}
  W_{\textrm{eff}} = C\left(\frac{\Lambda^{b_0}}{M_{ij}^a}\right)^{1/b},
\end{equation}
where $C$ is a normalization constant.  We have the condition that
\begin{equation}
  (b_0 - 2a)/b = 3,
\end{equation}
since $M_{ij}$ has mass dimension $2$.

Taking a derivative with respect to $M_{ij}$,
\begin{align}
  \frac{\partial W}{\partial M_{ij}} = 0 &= -C\frac{a}{b}\left(\frac{\Lambda^{b_0}}{M_{ij}^a}\right)^{(1 - b)/b}\frac{\Lambda^{b_0}}{M_{ij}^{a+1}} + yS_{ij}\\
  0 &= -C\frac{a}{b}\frac{\Lambda^{b_0/b}}{M_{ij}^{\frac{a}{b} +1}} + yS_{ij}\\
  \Rightarrow M_{ij} &= \left(\frac{C\frac{a}{b}
      \Lambda^{b_0/b}}{yS_{ij}}\right)^{1/(\frac{a}{b} + 1)}.
\end{align}
We'll assume that at the minimum we have solutions of the form $M_{ij}
= v^2\delta_{ij}, S_{ij} = s\delta_{ij}$.  Then the final equation
above is an equation for $v^2$ in terms of $s$.  While a derivative
with respect to $S_{ij}$ is
\begin{equation}
  \frac{\partial W}{\partial S_{ij}} = 0 = yM_{ij} + \gamma (S^2)_{ij}.
\end{equation}
Plugging in $M_{ij}$ in terms of $S_{ij}$ from above and evaluating
everything at the minimum (in terms of $v$ and $s$),
\begin{align}
  0 &= y\left(\frac{C\frac{a}{b}
      \Lambda^{b_0/b}}{ys}\right)^{1/(\frac{a}{b} + 1)} + \gamma s^2\\
  0 &= y\left(\frac{Ca}{yb} \Lambda^{b_0/b}\right)^{1/(\frac{a}{b} + 1)}s^{\frac{-1}{(\frac{a}{b} + 1)} - 2} + \gamma\\
  -\frac{\gamma}{y}\left(\frac{Ca}{yb} \Lambda^{b_0/b}\right)^{-1/(\frac{a}{b} + 1)} &= s^{-(2\frac{a}{b} + 3)/(\frac{a}{b} + 1)}\\
  s &= \left(\frac{-\gamma}{y}\right)^{-(\frac{a}{b} + 1)/(2\frac{a}{b} + 3)}
  \left(\frac{Ca}{yb}\Lambda^{b_0/b}\right)^{1/(2\frac{a}{b} + 3)}.
\end{align}
Using the mass dimension constraint to write $(2a/b) + 3 = b_0/b$,
\begin{equation}
  s =  \left[\frac{-y^a}{\gamma^{a+b}}\left(C\frac{a}{b}\right)^b\right]^{1/b_0} \Lambda.
\end{equation}
Therefore, there are potentially $b_0$ solutions and supersymmetric vacua.

The above analysis seems to be limited to the region of a theory with
an effective superpotential, like $SU(N_c)$ with $N_f < N_c$ which we
will explore in more detail below.  However, we shall see several ways
in which the same result for the number of supersymmetric vacua,
$b_0$, holds in regions where there is not an effective
superpotential.  For instance, one can work in the limit of very heavy
quarks, and integrate them out.  We will show this explicitly for
$SU(N_c)$, and this is again a very general procedure (although we
will not do this for a general theory).

\section{$SU(N_c)$ Models}\label{sec:su}

\subsection{$N_f < N_c$}
Working with $SU(N_c)$ supersymmetric gauge theory with $N_f < N_c$
flavors and $b_0 = 3N_c - N_f$, there is an effective superpotential
\cite{ads1, *ads2} given by
\begin{equation}
  W_{\textrm{eff}} = (N_c - N_f)\left(\frac{\Lambda^{b_0}}{\det M}\right)^{1/(N_c - N_f)}.
\end{equation}
Adding in the $N_f^2$ gauge singlet superfields $S_{ij}$, the
superpotential is now
\begin{equation}
  W = yS_{ff'}M_{ff'} + \frac{\gamma}{3}\Tr S^3 + W_{\textrm{eff}}.
\end{equation}
For simplicity, the singlet-quark couplings are all the same here, but
the features below are stable with small changes to the couplings.
One could also do a field redefinition.  We also take $yS$ such that
the quarks have a mass less than $\Lambda$.  We look for
flavor-symmetric solutions with all the $Q$'s having the same vev,
$v$, and $M_{ff'} = v^2\delta_{ff'}$.  Similarly, $S_{ff'} =
s\delta_{ff'}$.  Then $\det M = (v^2)^{N_f}$ at this point.  Taking a
derivative here,
\begin{align}
  \frac{\partial W_{\textrm{eff}}}{\partial M_{ff'}} &= -\Lambda^{b_0/(N_c - N_f)}(\det M)^{-\frac{1}{N_c - N_f} - 1}\frac{\partial \det M}{\partial M_{ff'}}\\
  &= -\Lambda^{b_0/(N_c - N_f)}v^{-2N_f(N_c - N_f+ 1)/(N_c -
    N_f)}(v^2)^{N_f - 2}\\
  \frac{\partial W_{\textrm{eff}}}{\partial M_{ff'}} &=
  -\Lambda^{b_0/(N_c - N_f)}v^{\frac{-2N_c}{N_c - N_f}}\delta_{ff'},
\end{align}
where we have used that $\partial \det A/\partial A_{ij} =
(A^{-1})_{ji}\det A$.  Then we have that
\begin{equation}
  \frac{\partial W}{\partial M_{ff'}} = ys\delta_{ff'} - \Lambda^{b_0/(N_c - N_f)} v^{\frac{-2N_c}{N_c - N_f}}\delta_{ff'},
\end{equation}
and setting this equal to zero and solving for $v^2$ (explicitly
putting in the phase),
\begin{equation}
  v^2 = \Lambda^{b_0/N_c}\left(\frac{e^{2\pi i k}}{ys}\right)^{(N_c - N_f)/N_c}.
\end{equation}
Taking a derivative of the superpotential with respect to $S_{ij}$ and
setting this equal to zero\footnote{Also using $\partial \Tr
  S^3/\partial S_{ff'} = 3 (S^2)_{f'f}$}
\begin{equation}
  \frac{\partial W}{\partial S_{ff'}} = 0 = yM_{ff'} + (S^2)_{f'f}.
\end{equation}
Working at the flavor-symmetric minimum, plugging in $M_{ff'} =
v^2\delta_{ff'}$ from above, and using that $S_{ff'} = s\delta_{ff'}$,
\begin{align}
  0 &= y \Lambda^{b_0/N_c}\left(\frac{e^{2\pi i k}}{ys}\right)^{(N_c - N_f)/N_c} + \gamma s^2\\
  0 &= \frac{y^{N_f/N_c}}{\gamma}\Lambda^{b_0/N_c}e^{2\pi i k (N_c - N_f)/N_c}s^{-(3N_c - N_f)/N_c} + 1\\
  \Rightarrow s &= \left(\frac{y^{N_f}e^{2\pi i k(N_c -
        N_f)}}{(-\gamma)^{N_c}}\right)^{1/(3N_c - N_f)}\Lambda.
\end{align}
Since $N_c$ and $N_f$ are integers, this implies $3N_c - N_f = b_0$
solutions.  This matches the general calculation in the previous
section. 

We note that when $\gamma \ll y$, $s$ is very large and thus the
quarks are heavy while the singlets are lighter.  In the opposite
limit, all the fields are much lighter.  We can also rewrite $v^2$
just in terms of the constants of the theory:
\begin{equation}
  v^2 = \left(\frac{-\gamma e^{4\pi i k}}{y^3}\right)^{\frac{N_c - N_f}{3N_c - N_f}}\Lambda^2.
\end{equation}

\subsection{$N_f \ge N_c$:}
There are (at least) two ways we can proceed to analyze the case $N_f \ge
N_c$: we can use the electric-magnetic duality or make all the flavors
heavy and integrate them out.  Let's start with the latter.

The concept of ``holomorphic decoupling'' (see, for instance, the
review \cite{dualityreview}) allows one to get the superpotential of
the theory from a known one of a theory with more flavors.  By making
these extra flavors heavy and integrating them out, one should
properly recover the behavior of the theory with fewer flavors.  This
is then also a constraint on the theory with more flavors, as it needs
to properly describe theories with fewer flavors in the decoupling
limit.  In the models we are considering here, the singlet
interactions always provide a mass term for the quarks.  If these
masses are made heavy by taking the singlet vevs to be large (compared
to the dynamical scale of the theory), the theory should have a
superpotential analogous to the case studied previously, with $N_f <
N_c$.

For these values of $N_f$, there are now also ``baryons'':
\begin{equation}
  B_{i_1i_2\cdots i_{N_f - N_c}} = \epsilon_{i_1i_2\cdots i_{N_f - N_c}  j_1j_2\cdots j_{N_c}} \epsilon_{k_1k_2\cdots  k_{N_c}}Q_{j_1k_1}Q_{j_2k_2}\cdots Q_{j_{N_c}k_{N_c}}
\end{equation}
where the $j$'s are flavor indices and the $k$'s are color indices.
There is a similar definition for ``anti-baryons,'' $\bbar{B}$, where
the $Q$'s are $\bbar{Q}$'s.  The baryons give new flat directions, and
we will add additional singlets to lift these as well.

Additionally, when $N_f = N_c$, the classical constraint of $\det M =
B\bbar{B}$ is modified to be $\det M - B\bbar{B} = \Lambda^{b_0}$ by
non-perturbative effects \cite{Seiberg:1994bz}.  There is no
superpotential generated.  This constraint can be implemented through
a Lagrange multiplier field.

We take all the fields but the singlets to be very heavy
(i.e.~$\gamma$ is small so the quarks get a heavier mass than the
singlets from the singlet vev), and assume that $\bbar{B} = B = 0$ at
the minimum.  Integrating out all the heavy degrees of freedom at the
scale $s$ (the singlet vev), the effective superpotential is
\begin{equation}
  W_{\mathrm{eff}} = \braket{\lambda\lambda} = s^3e^{-\frac{8\pi^2}{N_c g^2(s)}},
\end{equation}
where the denominator of the exponential has an $N_c$ because the beta
function coefficient is now for a pure gauge theory, $3N_c$, and the
$3$ cancels due to cubing the scale.  We have that
\begin{equation}
  \frac{8\pi^2}{g^2(s)} = \frac{8\pi^2}{g^2(\mu)} + b_0\ln\left(\frac{s}{\mu}\right),
\end{equation}
and $b_0$ is the coefficient of the beta function of the theory with
the massive quarks ($=3N_c - N_f$ for $SU(N_c)$).  Substituting this
in and rewriting,
\begin{equation}
  W_{\mathrm{eff}} = e^{-\frac{8\pi^2}{N_c g^2(\mu)}}\mu^{b_0/N_c}s^{3-\frac{b_0}{N_c}}.
\end{equation}
Again, after including an additional $s^3$ interaction term,
minimizing $W$ easily yields $b_0 (= 3N_c - N_f)$ solutions.  This is
a very general procedure, and can be used in the other theories we
consider as a way of extending the calculation of the number of vacua
beyond the region of an effective superpotential.

To enforce that the baryons are zero at the minimum, we can use
additional singlets, $\chi$ and $\tilde{\chi}$, one for each of the
$B$s and $\bbar{B}$s.  The additional terms in the superpotential are
then proportional to $\chi B$ and $\tilde{\chi}\bbar{B}$ (with all
indices suppressed).  The partial derivatives with respect to the new
singlets and baryons enforces that both are at zero.  These couplings
also need to be large enough to prevent a runaway in this direction.
In general the R-symmetry restricts any further terms with the new
singlets, but in some cases (due to the specific R-charges of the
theory) it may be necessary to impose some other symmetry as well.

Now let's see this using the electric-magnetic duality
\cite{seibergdual}.  The duality takes a theory at strong coupling to
one at weak coupling, and vice versa.  So here we are not in the same
coupling region as above, but we can consider our original theory at
strong coupling for a range of flavors of light quarks (relative to
$\Lambda$), and study it's analog at weak coupling through the
duality.  Again, there are a few ways to proceed here, but we will use
the duality to make a direct connection with the calculation for $N_f
< N_c$.  We will do this by showing that the effective superpotential
can be extended to larger values of $N_f$ by including the singlets.

The effective superpotential we studied above is not valid for $N_f
\ge N_c$, and the theory can instead be studied in its dual
``magnetic'' description \cite{seibergdual} (besides
\cite{dualityreview} another good review is \cite{isreview}).  Below
is basically a summary of some material in e.g.~\cite{isreview}; this
is a well known way to extend the previous results to larger $N_f$.

The magnetic gauge group is $SU(N_f - N_c)$ (matching the number of
indices of the baryon operators), with $N_f$ flavors of quarks $q_i$
and $\bar{q}_i$ and $N_{ij}$, a gauge invariant field.  The
superpotential for the dual theory is
\begin{equation}
  W = \frac{1}{\mu}qM\bar{q}.
\end{equation}
The scale $\mu$ relates the $M$ of the magnetic theory, $M_m$ (which
we will not use explicitly), with the $M$ of the electric theory,
$M_{ij} = Q_i\bbar{Q}_j$: $M = \mu M_m$.  The scale of the magnetic
theory is related to the electric theory by
\begin{equation}
  \Lambda^{b_0}\Lambda_m^{b_{m0}} = (-1)^{N_f - N_c}\mu^{N_f},
\end{equation}
where the ``m'' subscript denotes the magnetic theory, and $b_{m0} =
3(N_f - N_c) - N_f$ is $b_0$ for the magnetic theory.

In this dual picture, let's consider arbitrary values of $\braket{M}$,
so the magnetic quarks are massive, with a mass $\braket{M}/\mu$.  Now
the low energy theory has no matter besides the singlets, and the new
scale of the theory is
\begin{equation}
  \Lambda_{Lm}^{b_{Lm0}} = \frac{\det M}{\mu^{N_f}}\Lambda_{m}^{b_{m0}},
\end{equation}
where the low energy theory has beta function coefficient $b_{Lm0} =
3(N_f - N_c)$.

Gaugino condensation again leads to an effective superpotential:
\begin{equation}
  W_{eff} = (N_f - N_c)\Lambda_{Lm}^3 = (N_c - N_f)\left(\frac{\Lambda^{b_0}}{\det M}\right)^{1/(N_c - N_f)}.
\end{equation}
This is exactly the effective superpotential for $SU(N_c)$ with $N_f <
N_c$ flavors of quarks that we analyzed earlier, continued to this
value of $N_f$.

\section{$SO(N_c), Sp(2N_c),$ and $G_2$ Models}\label{sec:sosp}

Supersymmetric $SO(N_c)$ theories exhibit a very rich set of phenomena
\cite{son}.  In particular, aside from some special cases, there is a
dynamically generated effective superpotential which fits into the
general form of eq.~\eqref{eq:weff_gen}.  This is generated by gaugino
condensation as well, when $N_f < N_c - 4$ (and in some branch of the
theory when $N_c - 4 \le N_f < N_c - 2$) and $N_c \ge 4$, where $N_f$
is the number of flavors of quarks in the vector representation:
\begin{equation}
  W_{\mathrm{eff}} = A\left(\frac{\Lambda^{b_0}}{\det M}\right)^{1/(N_c - N_f - 2)},
\end{equation}
where $M_{ij} = Q_i Q_j$, $b_0 = 3(N_c - 2) - N_f$, and $A$ is a
normalization constant.  The anomaly coefficients are $2(N_c - 2)$ for
the adjoint and $2$ for the fundamental representations.

Since the effective superpotential for the $SO(N_c)$ theory is of the
same form as in the generic calculations of Sec.~\ref{sec:gen}, we
will again have a $\mathbb{Z}_{2b_0R}$ discrete R-symmetry and $b_0$
supersymmetric vacua.  For larger $N_f$, we can again integrate out
all the quarks (made heavy by the singlet vevs), or use a magnetic
duality to a $SO(N_f - N_c + 4)$ \cite{son}, similar to the $SU(N_c)$
calculations previously.

However, there are also several special cases for the $SO(N_c)$
theories.  When $N_f = N_c - 4$, the theory is broken to $SO(4) =
SU(2) \times SU(2)$, and so there are two gaugino condensates.  Only
when the condensates have the same relative sign does the theory have
the effective superpotential above.  On the other branch of the theory
there is no dynamically generated superpotential; there is a moduli
space, which includes confinement without chiral symmetry breaking
\cite{son}.  When $N_f = N_c - 3$, there is also a branch of the
theory which includes the effective superpotential.  Additionally,
when $N_c = 3, 4$ there are other considerations \cite{son}.

$Sp(2N_c)$ theories also fit easily into the general framework of
Sec.~\ref{sec:gen}, and are a bit simpler.  Here we have $2N_f$
flavors of quarks in the fundamental representation.  The anomaly
coefficients are $4(N_c + 2)$ for the adjoint and $2$ for the
fundamental representations.  For $N_f \le N_c$ there is a dynamically
generated superpotential (from gaugino condensation or through
instantons) \cite{spn}:
\begin{equation}
  W_{\mathrm{eff}} = A\left(\frac{\Lambda^{b_0}}{\mathrm{Pf} M}\right)^{1/(N_c
    + 1 - N_f)},
\end{equation}
where $b_0 = 3(2N_c + 2) - 2N_f$, $M_{ij} = Q_i Q_j$ is an
antisymmetric tensor (e.g.~the $Q$'s are combined with the
antisymmetric tensor that the group preserves), and $A$ is a
normalization constant.  Again, for larger $N_f$, the theory can be
analyzed by integrating out heavy quarks, or using a duality to a
$Sp(2(N_f - N_c - 2))$ theory.  There are no baryons for the
$Sp(2N_c)$ theories, as they break up into mesons by virtue of the
$\epsilon$ tensor being expressible in terms of the antisymmetric
tensor preserved by $Sp(2N_c)$.

The exceptional group $G_2$ also fits into this analysis quite easily
\cite{g2,exceptional}.  The beta function coefficient is $b_0 = 12 -
N_f$ and there are $N_f$ flavors of quarks in the fundamental
$\mathbf{7}$ representation.  There are several gauge invariant
fields; $M$ denotes the dimension two composite superfield.  The
effective superpotential is
\begin{equation}
  W_{\mathrm{eff}} = A\left(\frac{\Lambda^{b_0}}{\det M}\right)^{1/(4-N_f)},
\end{equation}
which again matches our general form.  This superpotential is
generated by gluino condensation for $N_f \le 2$ and by instantons for
$N_f = 3$.  For larger $N_f$ there is again a quantum modified moduli
space and then a dual picture for $N_f \ge 6$.

The exceptional groups present some difficulties in attempting to
extend this analysis, which is already apparent in the $G_2$ theory
\cite{exceptional}.  The more complicated group structure gives rise
to many gauge invariant composite fields, and so the effective
potential form cannot be completely fixed from general considerations.
Even so, the same arguments (such as the R-symmetry and flavor
symmetry) that give rise to the effective superpotential of
e.g.~$SU(N_c)$ are very general, and could possibly give the effective
superpotential for at least some region of the theories with other
exceptional groups \cite{exceptional}.  One way to do this is to
consider subgroups, which are reached by vevs of the different gauge
invariant composite fields.  For instance, the $\mathbf{27}$ of $E_6$ can break
the group to $SO(10)$ with a singlet, fundamental, and spinor
representation ($\mathbf{1} + \mathbf{10} + \mathbf{16}$).  This can have a generated
superpotential, and turning on the vevs will break this into smaller
and smaller subgroups.  So it seems quite possible that these results
could hold, with some restrictions, for theories with any simple Lie
group.

\section{Applications to Model Building}\label{sec:model}

We will now briefly discuss our results in the context of building
supersymmetric models, in a similar spirit to \cite{disr_paper}.  As
in \cite{disr_paper}, we have presented a mechanism for incorporating
a discrete R-symmetry and retrofitting (generating) mass scales of a
model by using the vev of a singlet field.  The overall goal of this
process is to make a model more ``natural:'' a mass hierarchy from
marginal or irrelevant couplings rather than imposed by hand.  The
R-symmetry can also be used to forbid unwanted operators.

First, let us consider how to build general models.  As a general
method, take the superpotential of the model to be modified and
replace masses by an interaction with a singlet field.  One may need
more than one singlet in order to generate different scales, or use
different coupling constants at the expense of some tuning or imposing
a hierarchy.  Additionally, one can choose the group, number of colors,
and number of flavors for the gauge theory the singlet(s) are coupled
to in order to produce the discrete R-symmetry desired.  The R-charges
of the model are now (at least partially) fixed due to the singlet
interactions.  This can forbid unwanted operators.

As an example, we consider a generalization of the Next to Minimal
Supersymmetric Standard Model (MSSM plus a singlet) as in the recent
work \cite{Delgado:2010uj} (see also the earlier work
\cite{Dine:2007xi}).  The superpotential of the Higgs and singlet
superfields is
\begin{equation}
W = \left(\mu + \lambda S\right)H_uH_d + \frac{1}{2}\mu_sS^2,
\end{equation}
where the cubic and linear terms for $S$ are assumed to be negligible
or set to zero.  This does not solve the $\mu$ problem, and in fact
adds another, $\mu_s$.  However, this model closely resembles the MSSM
phenomenologically and without tuning the scalar potential can have
the lightest neutral Higgs mass above current bounds and light top
squarks.

At first glance, this model seems rather unnatural: there are two free
mass parameters and some unwanted terms in the superpotential are
simply set to zero.  However, it is quite simple in our framework to
alleviate these problems.  We can use just a single additional
singlet, $\widetilde{S}$, coupled to this model to make it more natural, and
only require slight tuning of coupling constants to get any desired
hierarchy between $\mu$ and $\mu_s$:
\begin{equation}
W = \left(\frac{\alpha\widetilde{S}^2}{M_p} + \lambda S\right)H_uH_d + \beta\frac{1}{2}\frac{\widetilde{S}^2}{M_p}S^2.
\end{equation}
In order to have all these terms have R-charge $2$, the R-charge of the
Higgs and $S$ must be the same and $S$ has twice the R-charge of
$\widetilde{S}$ (all mod $2b_0$).  Then a cubic and linear term of $S$
are forbidden.  In \cite{Delgado:2010uj}, $\mu$ and $\mu_s$ are free
parameters taken at or below the TeV scale; $\mu = 500$ GeV and $\mu_s
= 2$ TeV are used in the plots.  Here, one would require $\beta
\approx 4\alpha$ to generate these relative scales.

\section{Discussion and Conclusion}

We have constructed models with a discrete R-symmetry that is
respected by the instanton, but gaugino condensation and the singlet
vev ultimately break it down to just a $\mathbb{Z}_{2R}$.  In
\cite{disr_paper} it was argued that this is not really an R-symmetry;
combined with a $2\pi$ rotation from the Lorentz group, this is just a
non-R $\mathbb{Z}_2$.  However, as discussed in \cite{disr_paper},
models with a discrete R-symmetry (larger than $\mathbb{Z}_{2R}$) can
be very important in models with low energy supersymmetry.

Due to the gaugino condensation and singlet vev, we showed that there
are $b_0$ vacuum states generally, and then in detail for the
$SU(N_c)$ theories.  This is basically a generalization of the
discrete $\mathbb{Z}_{2N_c}$ symmetry and resulting states in
$SU(N_c)$ with $N_f < N_c$ flavor of quarks and gaugino condensation.
We have analyzed general theories with singlet interactions to
construct the discrete R-symmetry.  The discrete R-symmetry was found
by considering an instanton and finding what discrete subgroup of the
$U(1)_R$ it respects: $\mathbb{Z}_{2b_0R}$.

We have analyzed many of the more common vector-like supersymmetric
gauge theories with an effective superpotential generated by gaugino
condensation or instanton effects.  We studied these with some
simplifications of the couplings.  Although these different theories
share many common features, in different regions of couplings there
are different behaviors.  The common form of the effective
superpotential for these gauge theories motivated a general
expression, which we used to show that we expect $b_0$ vacuum states.
For the known examples, the form of the effective superpotential
follows from renormalization, gauge symmetry, and the R-symmetry.
These are quite general arguments for any gauge group, but the exact
form of the effective superpotential is not completely fixed for the
exceptional groups.  These groups have several gauge invariant
composite fields and so there is still ambiguity in the form of the
effective superpotential.  On the other hand, our analysis of the
discrete R-symmetry, its breaking, and the number of vacua calculated
through integrating out heavy quarks still apply.  While a more
detailed study may be possible, our general results apply here as
well, thus all simple Lie groups.

There are still several open questions to pursue.  We did not
explicitly analyze the theories in the so-called ``conformal window,''
where the couplings and number of flavors would place the theory in a
conformal regime.  Although the previous analysis of integrating out
heavy quarks (where $\gamma$ is very small) still works, it would be
interesting to understand the dynamics of the theory in this regime.
How do the singlet interactions change the theory in this region?  Are
there general statements to be made here as well?  We have also
restricted our analysis to certain regions of the parameter space (and
largely ignoring baryons), and also limiting cases of integrating out
heavy matter fields.  Perhaps this can be made more precise, or maybe
there are interesting special cases to be found.  Theories like the
$SO(N_c)$ case also have more involved dynamics depending on the
number of flavors.  Again, maybe there is more to be said here as
well.  However, even at this point, the picture we have presented is
quite general and prevalent in supersymmetric Yang-Mills theories.

\begin{acknowledgments}
  The author would like to thank Michael Dine for not only proposing
  the topic studied in this paper, but also for many valuable
  discussions which ultimately resulted in this work, as well as
  helpful comments while preparing this manuscript.  Lorenzo Ubaldi
  also provided some comments on an earlier version of this
  manuscript.
\end{acknowledgments}

\renewcommand{\bibsection} {\section*{References}}
\bibliography{gen_gaugcond_ref}

\end{document}